# The Dark Components of the Universe Are Slowly Clarified

Vladimir Burdyuzha

Astro-Space Center, Lebedev Physical Institute, Russian Academy of Sciences,

Profsoyuznaya ul. 84/32, Moscow, 117997 Russia

Email:   burdyuzh@asc.rssi.ru

The dark sector of the Universe is beginning to be clarified step by step. If the dark energy is vacuum energy, then 123 orders are exactly reduced by ordinary physical processes. For many years these unexplained orders were called a crisis in physics. There was indeed a "crisis" before the introduction of the holographic principle and entropic force in physics. The vacuum energy was spent for the organization of new microstates during the entire life of the Universe, but in the initial period of its evolution the vacuum energy (78 orders) were reduced more effectively by the vacuum condensates produced by phase transitions, because the Universe lost the high symmetry during its expansion. Important problems of physics and cosmology can be solved if the quarks, leptons, and gauge bosons are composite particles. The dark matter, partially or all consisting of familon-type pseudo-Goldstone bosons with a mass of $10^{-5} - 10^{-3}$ eV, can be explained in the composite model. Three generations of elementary particles are absolutely necessary in this model. In addition, this model realizes three relativistic phase transitions in a medium of familons at different red shifts, forming a large-scale structure of dark matter that was "repeated" by baryons. We predict the detection of dark matter dynamics, the detection of familons as dark matter particles, and the development of spectroscopy for dark medium due to the probable presence of dark atoms in it. Other viewpoints on the dark components of the Universe are also discussed briefly.

# 1. Introduction

There has been some progress in understanding dark components of the Universe (these are: dark energy (DE) and dark matter (DM)).  The dark components account for 95% of the total density of the Universe. The baryons constitute only 5%. Our Universe is probably part of a perpetually growing fractal (multiverse). Some difficulties have appeared with the multiverse [1], and it is first necessary to examine our Universe, which probably tunneled from an oscillating regime to the Friedmann one at the moment of its birth [2]. The dark energy of the Universe is probably vacuum energy with the equation of state w≡ p/ρ = -1. The Planck experiment gives w= -1.006±0.045 for this ratio and $\Omega_{DE}$~ 0.69 for the dark energy density. In addition, the dark energy did not change in this experiment [3], which is surprising. Here, we emphasize that the vacuum of the Universe must evolve with time (the same conclusion was also reached in [4]). In our previous articles [5] we showed how as the crisis orders of dark energy are reduced if the dark energy is vacuum energy (for many years the lack of understanding of the reduction of 123 orders of dark energy was called a crisis of physics). We will discuss these calculations and experimental verification of dark energy models in more detail.

With regard to dark matter the situation is less clear ($\Omega_{DM}$~ 0.26 [3]), because it simply has no good explanation, although there are strong reasons to believe that of the dark matter consists of fimilons [6]. Phase transitions at different temperatures must have occurred in the medium of familons produced by quarks and leptons of all generations, distinguishing various scales of a fractal nature and forming a large-scale structure of dark matter. Baryons could repeat this picture owing to gravity, forming galaxies, clusters of galaxies, and even super clusters. We propose a composite model of elementary particles where the leptons and quarks are not fundamental particles.

Atoms of dark particles can probably also be formed. Estimates for the energy of some dark hydrogen and dark positronium ($e^+e^-$) transitions will be made. We will also mention the dark magnetic atom, monopolium ($g^+g^-$). Other dark matter models and their experimental verifications will also be discussed.

Note, for clarity, that $\Omega = \rho_i/\rho_{cr}$ is the dimensionless density of the component in the Universe normalized to the critical density $\rho_{cr}$. In addition, dividing the evolution of the Universe into epochs in the standard cosmological model can be of interest:

$$BB \rightarrow dS \rightarrow RD \rightarrow MD \rightarrow dS \rightarrow ......$$

Here, BB is the moment of its birth (T ~ $10^{32}$ K) followed by de Sitter stage (inflation), then the radiation-dominated epoch (RD), and then the matter-dominated epoch (MD). At present, there is an accelerated expansion (again the Sitter stage).

Each epoch was accompanied by important physical processes. After the birth, probably into a symmetric state, there was a quantum regime of evolution followed by inflation (the generation of initial perturbations during it [7], phase transitions, reheating, particle production, nucleosynthesis, etc. The Universe cooled down, and there was recombination at a red shift z~ 1100. At this moment (the end of the radiation-dominated epoch) the Universe cleared up, and its age was then only 400 000 years. The picture of microwave sky gives us the brightness distribution from the sphere of last scattering. At present (z=0), the temperature of the microwave sky is T ~ $2.7^0$ K, i.e., the Universe cooled down to such temperature.

After recombination there was the matter-dominated epoch. During the matter-dominated epoch (this is practically the entire age of the Universe, 13.8 x$10^9$ years) all of the observable baryonic structures and dark matter structures, about which nothing is known as yet, developed from initial perturbations.

The Universe expanded with a deceleration up to a red shift z~0.7 due to the influence of gravity, and it has expanded with an acceleration, i.e., again the de Sitter stage, already for almost $7 \times 10^9$ years. The subsequent evolution of the Universe is equally interesting, but we will not discuss it here. Note only that the Universe during the matter-dominated epoch also had a period of secondary ionization, when quasars light up (z~11).

But let us return to the dark components, the most recent on which can be found in [8, 9]. The probable DE and DM models that, in my opinion, are appropriate for cosmology are also discussed in this review.

## 2. Dark Energy

To avoid confusion, in this paper dark energy, vacuum energy, the cosmological constant, and Λ-term are identical concepts. The cosmological constant (Λ-term) problem existed for many years, because there was no understanding of its reduction by 123 orders of magnitude [10]. The problem was even called a crisis of physics. Indeed, this crisis existed before introduction of the holographic principle [11] and entropic force [12] in physics. The cosmological constant has been introduced by A. Einstein [13] in his field equations as a property of space to preserve a static Universe:

$$R_{\mu\nu} - (1/2) R g_{\mu\nu} + \Lambda g_{\mu\nu} = -8\pi G_N T_{\mu\nu} \qquad (1)$$

Einstein rewrote the rules of our physical world by this equation 100 years ago. However, not all was that smooth. A. Friedman [14] showed that the static solutions of the Einstein's equation (1) are unstable and even a minuscule matter density fluctuation leads to collapse or perpetual expansion. A. Einstein then abandoned the Λ-term, because it became known that the Universe is expanding.

Much later, in 1947, in Lemaitre's letter to Einstein[15] the idea of the cosmological constant seemed brilliant to Lemaitre, an abbot and a mathematician. He interpreted it as the vacuum energy density, i.e., the Λ-term is the vacuum energy of the Universe. If we move the Λ-term to the right-hand side of Einstein's equation (1), then this will be a form of energy that was called dark energy. We believe that the dark energy should be associated with the vacuum energy. A vacuum with the equation of state $p = -\rho$ is a stable state of quantum fields without any excitation of wave modes (nonwave modes are condensates). Consequently, the vacuum of the Universe consists of quantum field condensates that are diluted and, of course, fluctuate during its expansion. The energy density of the present vacuum at redshift $z = 0$ differs significantly from the energy density at the birth of the Universe, i.e., at $z = \infty$:

$$\rho_{DE} \sim 10^{-47} (GeV)^4 \text{ for } z=0 \text{ and } \rho_{DE} \sim 2\times 10^{76} (GeV)^4 \text{ for } z=\infty$$

Here, we have a difference in vacuum energy density by 123 orders of magnitude that were reduced through rather ordinary physical processes during the evolution of the Universe. Our Universe was probably born by tunneling from an oscillating regime to the Friedmann one [2] and began to expand. New microstates arise during its expansion (evolution). Fock's ideas [16] should be used to construct the space of new quantum states.

However, we can take a different path. An entropic force emerges as the Universe expands, with the energy losses being inevitable due to its presence. Only the vacuum energy can be the energy source. The entropic force *F* was proposed by E. Verlinde [12] as a microscopic force:

$$F \Delta x = T \Delta S \text{ or } F \sim T \partial N / \partial x \qquad (2)$$

where Δ*S* is the change in entropy at a displacement Δ*x*, and *N* is the information about the holographic system in bits. In cosmology the idea of an entropic force was already applied in [17] for the same "crisis" purposes, but unsuccessfully, although the very existence of an entropic force allows one to talk about the energy (*F*Δ*x*) that, in this case, is taken away from the vacuum energy as the Universe expands. Thus, the ideas of holography should be applied more properly, i.e., formally only in the classical regime. J. Maldacena [18] pointed out that AdS/CFT correspondence asserts all information about a gravitational system is coded at its boundary. In the holographic limit the vacuum energy density of the Universe is then [19] $\rho_{DE} \leq 3M_{pl}/8\pi R^2$, where $M_{Pl}$ is the Planck mass. In addition, J.Bekenstein [20] showed that the entropy (the number of microstates) of a black hole is 1/4 of the area of the event horizon expressed in Planck units. The idea of a similarity of the thermodynamics of a black hole in special coordinates to the thermodynamics of a de Sitter Universe belonging to S. Hawking [21] turned out to be very useful, as did T. Jacobson's idea [22] that gravity on the macroscopic scale is a manifestation of vacuum thermodynamics. Our Universe, after its creation [2], passed the quantum stage of its evolution, when holographic ideas can not used, since holography is a classical phenomenon. Probably, in the quantum regime our Universe lost its high symmetry, extra dimensions, and parity, but at the same time, a whole bunch of particles, including the dark ones, was produced. Of course, there was also a transitional regime between the quantum and classical regimes.

Upon sharp cooling after the birth of the Universe, when its high symmetry was lost, a number of phase transitions occurred. Their condensates compensated 78 orders of vacuum energy [5, 23], because the forming condensates made negative contributions to the positive vacuum energy density. There could the following chain of phase transitions in our Universe:

P→D$_4$x[SU(5)]$_{SUSY}$→D$_4$x[U(1)xSU(2)xSU(3)]$_{SUSY}$→D$_4$xU(1)xSU(2)xSU(3)→D$_4$xU(1)xSU(3)→D$_4$xU(1)

$10^{19}$ GeV      $10^{16}$ GeV           $10^5$~$10^{10}$ GeV         100 GeV         0.265 GeV

In this case, huge orders of vacuum energy were could be lost during quantum regime (~$10^{-6}$ sec). Of course, this chain could be more complex. Thus, phase transitions in the early Universe could quench more than 78 orders of the vacuum energy:

$\rho_{Pl}/\rho_{QCD}$ ~ $(M_{Pl}/M_{QCD})^4$ = $(1.22 \times 10^{19} / 0.265)^4$ ~ $4.5 \times 10^{78}$     (3)

The last two last phase transitions of them could quenched 10 orders of vacuum energy. Fortunately, the last two condensates of this chain can be calculated exactly within the Standard Model [5]. One of them is a Higgs condensate in the electro-weak interactions theory ($\rho_{EW}$); the other is a quark-gluon condensate in quantum chromodynamics ($\rho_{QCD}$). At the Higgs mass is $m_H$ ~ 125 GeV for a Higgs condensate we obtain:

$\rho_{EW}$=-$m_H^2 m_W^2/2g^2$-$(1/128\pi^2)(m_H^4+3m_Z^4+6m_W^4-12m_t^4)$ ~ -$(100 GeV)^4$.

(4)

For a quark – gluon condensate we have:

$\rho_{QCD}$=-$(b/32)<0|(\alpha_s/\pi)G_{ik}^a G^{ik}_a|0>$ ~ - $(265 MeV)^4$     (5)

Then, $\rho_{EW}/\rho_{QCD} = (100/0.265)^4 \sim 2 \times 10^{10}$. At the beginning of the last phase transition the Universe had a density of $\sim 10^{-2}$ (GeV)$^4$ or $10^{16}$ g cm$^{-3}$. Probably, the quantum regime of the Universe evolution took place during $10^{-6}$ sec and a minimal size of the Universe ($R_{QCD}$) at the begining of the classical regime could be near $3 \times 10^4$ cm. There should also be a transient regime between the quantum and classical ones. Of course, there is some uncertainty about application of the holographic principle here. If we take the beginning of the classical regime at t=$10^{-5}$ sec, then we can get an exact result of vacuum energy reduction.

By now (z=0) the vacuum energy must diminish further by a factor $\rho_{QCD}/\rho_{DE} \sim (0.265/1.8 \times 10^{-12})^4 \sim \mathbf{5 \times 10^{44}}$ to quench all 123 orders (in the quantum regime the Universe already lost 78 orders of vacuum energy, as we have already noted).

How can the vacuum energy losses by 44 orders of magnitude be obtained and what process is "guily" of this? We have a physical basic, the entropic force that emerges as the Universe expands, and S. Hawking's assertion about a similarity of the thermodynamics of a de Sitter universe to the thermodynamics of a black hole. In addition, C. Balazs and I. Szapidi [19] argueded that the entropy of the Universe is bounded by its "surface" measured in the Planck units: $S \leq \pi R^2 M_{Pl}^2$. This surface serves as a holographic screen. In the holographic limit the vacuum energy density of the Universe is then related to the entropy by a very simple formula: $\rho = 3 M_{Pl}^4 / 8S$ that for calculations in the classical regime is

$$\rho(z) [GeV]^4 = (3/8) M_{pl}^4 [R_{QCD} / R(z)]^2 \qquad (6)$$

At z=0 we have: ρ (0) = 0.375x10$^{-47}$(GeV)$^4$ if R(0)= 10$^{28}$ cm. In the classical regime of the Universe evolution the vacuum energy could be reduced by a factor of (3/8)(10$^{28}$/3x10$^4$)$^2$ ~4x10$^{46}$ in 4x10$^{17}$ sec. If the beginning of classical (Friedmann) evolution is taken at a size ~3x10$^5$ cm, then we will have a coincidence of the reduction in vacuum energy with the "required" value, i.e., (3/8()10$^{28}$/3x10$^5$)$^2$~**4x10$^{44}$**. Note that there is some arbitrariness in the estimates here, because we do not know how long the transitional regime was. On the other hand, this almost exact coincidence cannot be fortuitous.

In our early publications [5] about it we discussed an application of these approximations to cosmology. General relativity provides a bright example of the holographic theory, while the existence of a horizon in the Universe gives "a strong argument" for the holographic approximation in cosmology. The growth of the information entropy in the Universe during its expansion is obvious. The existence of a holographic limit [24] constrains the number of degrees of freedom (the number of microstates) that can exist in a bounded volume. Both sizes ($R_{QCD}$= 3x10$^5$ cm and R=10$^{28}$ cm), can be causal horizons in the holographic thermodynamics of the Universe. The Einstein's equations are obtained from proportionality of the entropy to the event horizon, given the Clausius fundamental relation dS=dQ/T, where dS is the change in entropy, dQ is the change in energy flow through the horizon, and T is the Unruh temperature seen by an accelerated observer inside the horizon [20]. In a de Sitter space, the event horizon coincides with the apparent horizon. Some cosmological models dispense with the event horizon, but the apparent horizon exists always.

In conclusion, note some facts related to the cosmological constant (this constant was a brave A. Einstein's ansatz).

1) Already at the Planck scale, the 3-dim topological defects (wormholes) of the gravitational vacuum condensate [5] diminished the positive initial Λ-term:

$$\Lambda_{QF} = \Lambda_0 - (\kappa \hbar^2 / 768\pi^2) c_3^2 \qquad (7)$$

here: $c_3$ is a constant and $\kappa = (10^{19})^{-2}$. In additional, this condensate could fix the beginning of time in our Universe [25].

2) The super symmetry is broken if and only if the cosmological constant is positive.

3) Many years ago Ya. Zel'dovich [26] attempted to find the vacuum energy of the Universe in terms of zero-point oscillations, using formula from fundamental constants derived by him:

$$\rho_\Lambda [g\ cm^{-3}] = G_N\ m^6 c^2 h^{-4}. \qquad (8)$$

His attempts were unsuccessful, because he substituted the electron or proton mass into this formula. However, if $\pi$-meson mass is substituted, then the situation changes. The chiral QCD symmetry is not an exact symmetry, and the appearance of pseudo-Goldstone bosons ($\pi$ –mesons in our case) is a physical manifestation of this symmetry breaking. If we substitute the average mass of $\pi$-mesons in Eq. (8) and take the Hubble constant to be $H_0$=70.5 (km sec$^{-1}$/Mpc), then we will obtain $\Omega_\Lambda = \rho_\Lambda / \rho_{cr} \sim 0.73$. Thus, the relative content of the vacuum component was fixed in a very early Universe. Note that the astrophysical parameters slightly "drift" from one experiment to another. From the results obtained by Planck satellite we have $H_0$=69 km s$^{-1}$ Mpc$^{-1}$ and $\Omega_\Lambda$~0.69, while we used the astrophysical parameters from the 7- year-long WMAP experiment [27] in our calculations of $\Omega_\Lambda$.

4) If gravity on the macroscopic scale is an entropic force [22], then gravity is not a fundamental interaction, which was pointed out by A. Sakharov [28] long ago, but, of course, he had other arguments.

5) The distance dependence of the vacuum energy (Eq.6) is a new point in the vacuum thermodynamics of the Universe. This is a nonstandard view, because the traditional vacuum energy with the equation of state w=-1 must not change. However, first, observations give w≠-1, though the difference is very small (see Introduction), and, second, the holographic approach and the entropic force based on which our concept of lifting of "anathema" from the cosmological constant was constructed are also far from the universality accepted physical standard.

Let us give several absolute values of the vacuum energy densities that follow from our calculations to show its evolution [5]: $\rho_\Lambda$~0.375x10$^{-47}$, $\rho_\Lambda$~31x10$^{-47}$, $\rho_\Lambda$~197x10$^{-47}$ at z=0, 5, and 10, respectively. These values are correct if and only if dark energy is a purely vacuum energy. Our calculations of the vacuum energy density up to the red shift z=10$^{11}$ (the age of the Universe was then only t=0,003 s) are presented in [5]. The red shift dependence of the vacuum energy density is roughly quadratic. A good review on the dark energy problem can be found in recent paper [29].

Finally, vacuum (dark) energy of our Universe has evolved from the Planck time until now. The Universe lost ~ 123 ($4 \times 10^{78}$ x $4 \times 10^{44}$) orders of this form of energy during $4 \times 10^{17}$ sec in the process of creating new microstates as it expanded (in the quantum regime the phase transitions were more effective in this reduction). Thus, the crisis of physics related to the cosmological constant that lasted for many decades can be overcome. As this problem was solved, a gravitational ether was introduced in [30] for a thermodynamic description of gravity, although Verlinde's entropic force [12], from our viewpoint, is more descriptive in cosmology. In [30] the crisis cosmological constant problem is already called old, because there are also problems in the new model with the gravitational ether.

There are also other dark energy models. The simplest candidate for DE [31] can be an extremely low-mass scalar field $\phi$ with effective potential $V(\phi)$. As the field $\phi$ decreases slowly, its constant potential energy can be responsible for the creation of late inflation, which is probably observed at present (though there is doubt about the existence of late inflation). The initial inflation, for which the scalar field is responsible, has almost nobody been doubted. Effects from a modification of general relativity [32] can give a theoretical explanation for the observed accelerated expansion of the Universe. In addition, more careful processing of the data from the Planck satellite (their third version [3]) showed that the equation of state for dark energy coincides with the data of the first version, i.e. w = -1.006 ±0.045. This suggests that the dark energy is not a pure vacuum energy but has a small admixture of, probably, a scalar field (for a discussion on this, see also [33].

Many groups are preparing experiments with DE. In the Archimedes experiment [34] (Italian group) the vacuum of the Universe will be weighed using the Casimir effect. Simultaneously several experiments on this subject matter are being prepared in USA [35], and the international DES consortium even exists (for reviews on DE problem, see [36]). Finally, a discussion of various models for the accelerated expansion of the Universe can be found in [37]. In addition, the author of [37] gives a new definition of the vacuum: this is a "fully coherent state including all Fock states of the Universe". We expect an intriguing decade for a complete clarification of the dark energy problem and, of course, a confirmation of its evolution.

## 3. Dark Matter

Dark matter particles are particles with a cross-section for their collisions smaller than the cross-section for the collision of baryons by 20-25 orders of magnitude. Back in the 1930s Swiss astrophysicist Fritz Zwicky, who worked at that time in the USA at a large telescope, discovered that the Coma cluster of galaxies could not be gravitationally bound without the existence of an additional mass. Subsequently, this mass was called dark matter. Much later, the existence of dark matter was confirmed from observations of rotation curves for galaxies and directly through gravitational lensing. Quite recently, it has become clear that the dark matter in the Universe is more abundant ($\Omega_{DM}$~0.26) than visible (baryonic) matter ($\Omega_b$~0.05). A new branch appeared in cosmology, dark matter cosmology. The dark matter is probably a mixture of various particles, such as neutralinos, axions, right-handed neutrinos, and dark atoms with different percentage contributions to the total density. There are a lot of assumptions about the dark matter composition.

We would immediately like to present our dark matter model that will help to solve important questions in physics and cosmology. This is a composite model of elementary particles that incorporates familons (familons are a kind of axions) in the dark matter. Our model explains the necessity of the existence of three particles generations in the Standard model (SM), distinguished scales in the Universe, as well as the fractal distribution of material components. Note that the SM of particle physics is not a complete description of nature. It probably gives good results at energies up to tens of TeV. New physics will most likely be required to explain dark matter.

Observational data suggest that the large-scale baryonic structure was formed at red shift z~6-8 or even earlier [38]. For the SM in cosmology (ΛCDM) the existence of baryonic structures at these red shifts is difficulty, because there is very little time for the primordial perturbations to develop into observable structures against the background of an expanding Universe. If baryonic structures were formed at red shifts z >10, then a key role in such process must be played by dark matter particles, which could prepare a medium for the condensation of baryons.

In the standard cosmological model, one of the DM components can be a light particle with a mass of m~$10^{-5}$eV (axion), which barely interacts with particles of ordinary matter. The axions have not yet been discovered due to their super weak interaction with baryons and leptons. Dark matter particles (familons in our case) at a specific evolution stage of the Universe lost their residual U(1) symmetry and formed a large-scale structure of dark matter as a result of relativistic phase transitions, thereby preparing a medium for the condensation of baryons that subsequently became the large-scale (already baryonic) structure of the Universe. Only a critical phenomenon (phase transition) could create a fractal structure in the distribution of density perturbations in the dark matter.

Let us make several remarks explaining why we resort to the composite model. There are 12 particles (24 with antiparticles) in the SM of particle physics that form a repetitive picture, forcing us to assume that they are not fundamental particles but consist of smaller particles, preons. It is well known that the SM was not complete before the discovery of the Higgs boson at the Large Hadron Collider (LHC) [39, 40]. SM contains three columns showing three generations of particles. We know that the heavy quarks decay into the lightest ones (u, d), because the quarks in the second and third generations are heavier than those in the first one. The quarks of the third generation (t, b) are heavier than those in the second one (c, s). The picture for leptons is similar. The electron has heavy partners, the muon (μ) and the taon (τ), and there are also three flavors of neutrinos in the SM. Of course, we can talk about the decay of heavy leptons into light ones (μ⁻ →e⁻ + γ), but such decays have not yet been observed. The limits for the mass of all particles can be found in the latest publication of Particle Data Group [41]. A faint hint that elementary particles are composite objects can be the radioactive decay of nuclei. Therefore, it is interesting to consider the composite (preon) model in more detail. Following [42, 43], all elementary particles can be described in the preon version according to the table (its full version is contained in [6]). An example is presented in the table, where it is assumed that there are several types of preons: the + preon with electric charge + 1/3 and the 0 preon without electric charge; the − antipreon has electric charge -1/3 and the neutral antipreon is designated as □.

**Table 1.**

| Particle | Preon composition | Electric charge |
|---|---|---|
| **Positron** | +++ | +1 |
| **Down quark** | -□□ | −1/3 |
| **Up antiquark** | - -□ | −2/3 |
| **Electron antineutrino** | □□□ | 0 |
| **W⁺** | +++000 | +1 |

Formally, a subpreon model, where even the preons are composite particles, has already been proposed [44]. In addition, a natural question arises: why there are three generations of particles in the SM, when can the first generation be sufficient, and whose particles are the observable baryonic Universe? In the preon model the second and third generations of particles are excited states of the first- generation particles (the second- and third-generations particles are unstable). The second- and third-generations particles must then consist of combinations of the same preons and antipreons as the first generation. Of course, preons and antipreons are fermions. The same building blocks could form gluons. Preons have not been observed experimentally, although attempts have been made to observe them. First, the fourth generation of particles and, second, the rare decays of leptons $\mu^- \to e^- + \gamma$, along with some sophisticated experiments (for example, a measurement of the muon magnetic moment), can point to the internal structure of quarks, leptons, and gauge bosons. An upgrade of the LHC led to an increase in the proton beam energy and dramatically increased the probability of detecting events on smaller scales, which is required for the search of preons.

Let us return to our version of dark matter. The authors of [45,46] considered dark matter as a gas of pseudo-Goldstone bosons, in which there could be a phase transition at ultralow temperatures compared to the collider temperatures. Pseudo-Goldstone bosons (including familons) emerge when the continuous symmetry of vacuum is broken. The familon symmetry is a horizontal symmetry broken by a Higgs condensate [46]. The familon symmetry breaking manifests itself in the different masses of the particles of the three generations. An important fact for us is that familons possess a residual U(1) symmetry which takes place if some Goldstone degrees of freedom are not transferred to vector states. The properties of any pseudo-Goldstone bosons (there are four types of them), along with familon–type pseudo-Goldstone bosons, depend on the realization of the Goldstone modes. These modes can emerge both from the fundamental Higgs fields [46], and from collective excitations of a heterogeneous nonperturbative vacuum condensate, which, in our case, is more complex than the quark-gluon condensate. This is possible in theories where leptons, quarks, and gauge bosons are composite objects, i.e., in the preon model of elementary particles.

Consider the simplest boson-fermion-preon model of left chiral quarks and leptons. The basic elements of this model are the chiral fermion preons $U^\alpha_L$ $D^\alpha_L$ and the scalar preons of quark $\phi^{i\alpha}_a$ type and lepton $\chi^\alpha_l$ types. The internal structure of elementary particles in our model will then be

$$u^i_{La} = U^\alpha_L \, \phi^{i\alpha}_a \qquad u^i_{La} = (u^i_L, c^i_L, t^i_L)$$

$$d^i_{La} = D^\alpha_L \, \phi^{i\alpha}_a \qquad d^i_{La} = (d^i_L, s^i_L, b^i_L)$$

$$\nu^i_{Ll} = U^\alpha_L \, \chi^\alpha_l \qquad \nu^i_{Ll} = \nu_{Le}, \nu_{L\mu}, \nu_{L\tau}$$

$$l^i_{Ll} = D^\alpha_L \, \chi^\alpha_l \qquad l^i_{Ll} = (e_L, \mu_L, \tau_L)$$

(9)

In the case of a leptoquark (a strongly interacting particle), our model gives

$$(LQ)_{al} = \phi^{i\alpha}_a \, \chi^\alpha_l \tag{10}$$

Note that the leptoquarks (LQ) are scalar or vector (so far hypothetical) bosons carring both baryon (B) and (L) numbers. In Eqs. (9), (10) i is the color index of quantum chromodynamics, a, b, c = 1, 2, 3; l, m, r = 1, 2, 3 is the number of quark and lepton generations, and $\alpha$ is the metacolor index, corresponding to the new metachromodynamic interaction that binds preons into quarks and leptons. Inside quarks and leptons, the metagluonic fields $G^{\omega}_{\mu\nu}$ and the scalar preon fields $\phi^{i\alpha}_a$ and $\chi^{\alpha}_l$ are in the state of confinement. This effect is similar in its physical nature to the confinement of quarks and gluons inside hadrons, providing the existence of nonperturbative metagluonic and preon condensates. These condensates are described by the following relations:

$$<0| (\alpha_{mc}/\pi) G^{\omega}_{\mu\nu} G^{\mu\nu}_{\omega}|0> \sim \Lambda_{mc}^4 \quad (11)$$
$$<0| \phi^{i\alpha}_a \phi^{i\alpha}_b |0> = V_{ab} \sim -\Lambda_{mc}^2 \quad (12)$$
$$<0| \chi^{\alpha}_l \chi^{\alpha}_m |0> = V_{lm} \sim -\Lambda_{mc}^2 \quad (13)$$

Here $\Lambda_{mc}$ is the energy scale of preon confinement, $V_{ab}, V_{lm}$ are the condensate matrices. The condensates (11) and (12), together with the gluonic condensates $<0|(\alpha_c/\pi)G^a_{\mu\nu} G^{\mu\nu}_a|0>$ and the quark condensates $<0|\bar{q}_L q_R + \bar{q}_R q_L|0>$ provide a quark mass creation mechanism for all three particle generations. In [47], we illustrated the mass creation mechanism for all three particle generations by a special figure and discussed the structure of the condensate matrices in detail.

Within the framework of this theory, DM is a system of familon collective excitations of a heterogeneous nonperturbative vacuum consisting of three subsystems: up-quark-type familons; down-quark-type familons; lepton-type familons. The small familon masses of are the result of superweak interactions of Goldstone fields with nonperturbative vacuum condensates. These masses are constrained by the astrophysical and laboratory data: $m_{astr} \sim 10^{-3}$-$10^{-5}$ eV; $m_{lab} < 10$ eV.

At stages of the cosmological evolution which are far from the quarkonization and leptogenesis (T«$\Lambda_{mc}$) there are already no heavy unstable familons . The fate of each subsystem of the low-energy familons depends crucially on the sign of the square of the rest mass created by the interaction of familons with quark condensates.

Formally the familon mass production effect corresponds to the appearance of mass terms in the Lagrangian of Goldstone fields. From general considerations we may assume that mass terms can arise with both "correct" and with "incorrect" signs. The sign of the mass terms predetermines the destiny of residual symmetry of Goldstone fields. In the case of an "incorrect" sign, at low temperatures $T<T_c \sim m_{familons} \sim 0.1 - 10^5$ K, a Goldstone condensate is produced when the familon gas symmetry is spontaneously broken. As shown in our papers [6, 47], for the complex scalar field the square of the masses was negative for both up-and down-quark –familon subsystems:

$$m_{f(u)}^2 = -(1/24u^2) <0 | (\alpha_s/\pi)G_{\mu\nu}^{\ n} G^{\mu\nu}_{\ n} | 0> [(m_t - m_c)^2/m_c m_t]$$

$$m_{f(d)}^2 = -(1/24u^2) <0 | (\alpha_s/\pi)G_{\mu\nu}^{\ n} G^{\mu\nu}_{\ n} | 0> [(m_b - m_s)^2/m_b m_s]$$

(14)

Here, u is the average value of the scalar field in the condensate. This means that pseudo-Goldstone vacuum is unstable at $T< T_c < |m_f|$. A relativistic phase transition to a state with spontaneous breaken U(1) symmetry must occur in gas of psevdo-Goldstone bosons at $T=T_c$. The subsystem of leptonic familons is also unstable and a relativistic phase transition occurs in it, but here we face an unstudied leptonic condensate (or possibly a lepton – quark condensate).

The phase transition in the cosmological familon gas is a first – order phase transition with a wide domain of phase coexistence. Numerical simulations of such relativistic phase transition have shown that there was an alternation of high-symmetry and low-symmetric phases in the Universe with a density contrast $\delta\rho/\rho$ ~0.1. In the period of phase coexistence [47], the characteristic scale of such block-phase structure is determined by the horizon size at the time of the relativistic phase transition, i.e., in other words, a large-scale structure of dark matter can be formed. Since there are three familon subsystems, our model makes it possible to realize three relativistic phase transitions.

Consequently, this part of the dark matter consisting of pseudo-Goldstone bosons of familon type is a multicomponent heterogeneous system evolving in a complicated thermodynamic way. It is composed of nine particles with different masses. During its evolution this system underwent three relativistic phase transitions that took place at different temperatures. Of course, the thermodynamic temperature of the familon gas may not coincide with the thermodynamic temperature of other subsystems of the Universe. At the present epoch this can manifest itself in the fact that the temperature of the familon gas (as a part of the dark matter) can differ from the temperature of the cosmic microwave background.

Thus, the proposed model is unambiguously connected with the preon model of elementary particles which has good prospects for experimental verification at colliders. There is no doubt that an experimental status can be given to our model only after the discovery of familons. Authors [48] already discussed a possible search for composite particles at the LHC before its upgrade by studying the dimuon mode in pp collisions at the Compact Muon Solenoid (CMS). After acquisition of an experimental status, our model will be inevitable accepted in cosmology, because not only the role of particle generations but also their number is clarified. At least three generations of particles are needed for the structuring of the Universe to be possible: the first generation of particles gives the observable baryon world, the second and third generations (their existence) give all or part of the dark matter. In this model the second and third generations are excitations of the first generation. Only in the preon model of elementary particles will the fractality in the distribution of material components of the Universe due to phase transitions be explained naturally. In the preon model the three distinguished scales of the Universe may also be explained naturally (galaxies, cluster of galaxies and super clusters of galaxies), because the phase transitions occurred at different red shifts.

Recall that axions emerged in a well- motivated extension of the SM to solve the long-standing strong-CP problem. They are also candidates for dark matter, along with weakly interacting massive particles (WIMPs) from super-symmetric theories (neutralino). Attempts were made at the University of California to solve the dark matter problem and the "White Paper" on the results of this brainstorming workshop was published [49]. In some theories the favorite is a mixed neutralino-axion model [50] (the author of this review also holds this viewpoint). Note that great efforts are being made to find the axions. The International Axion's Observatory [51], along with the Center for Axion and Precision physics [52, has already been created (recall once again that familons are a kind of axions).

Let us return to the composite model or, more precisely, to the probable detection of a scalar leptoquark at the LHC (see Eq.10)). Recently, the authors of [53] pointed out that the CMS collaboration observed anomalous events in *pp* collisions at √s=8 TeV that can be interpreted as the production of a scalar leptoquark (three –color boson). Of course, this highly interesting result has caused a barrage of papers on this subject, because the existence of leptoquarks is the simplest extension of the SM. They also appear in such extensions of the SM as the Grand Unification SU(5) [54], in technicolor models [55], and, as has already been noted, in composite models [56].

Here, we will restrict ourselves to the report of only the CMS collaboration itself [57] and to the paper [58] whose author investigated the possible detection of leptoquarks at the LHC. Note, however, that in [47] we also pointed out the probable detection of leptoquarks. The authors of [53] even estimated the mass of the leptoquark ($m_{LQ} \approx 650$ GeV) as a possible candidate for explaining the observed events in the CMS experiment [56] and discussed the models where a leptoquark decays into dark matter particles and leptonic jets. An important remark was made in [59]. The neutrinos with energy ~ $10^{15}$ eV detected in the Ice Cube experiment can test or can even be associated with the leptoquarks, which, of course, is intriguing (there is a reference to such neutrinos in the conclusion).

Interesting evidence for a possible trace of DM has recently been found [60-61]. The detected X-ray line (3.55 keV) does not coincide with any known spectral line of any baryonic atom. An axion-like particle (its decay in a magnetic field) can provide the simplest explanation for the observed line: it can be the decay of a right-handed neutrino into a left-handed neutrino and a gamma –ray photon [62]. However, if atomic dark matter can be realized, then it can naturally explain this line without using WIMP paradigm. For example, the 3.55-keV X-ray line can be the

hyperfine –splitting line of dark hydrogen [63], analog of the 21-cm line of the ground state of atomic hydrogen:

$$\Delta E = (8/3) (\alpha')^4 (m_e^2 m_p^2)/m_h^3 = 3.55 \text{ keV} \qquad (15)$$

Here $m_h = m_e + m_p$ and $m_h$, $m_e$, $m_p$ are the dark hydrogen, dark electron, and dark proton masses. The authors [63] suggest that the dark hydrogen mass $m_h$ can lie in the range 350-1300 GeV, the fine structure constant $\alpha'$ is in the range 0.1 - 0.6, $m_p/m_e = 10^2 - 10^4$. These quantities satisfy Eq. (15). Several interpretations of this X-ray line already exist [64]. If the 3.55 keV line is actually the hyperfine-splitting line of dark hydrogen, then it is natural to expect the detection of the $L_\alpha$ and $H_\alpha$ lines of dark hydrogen (along with other strong lines). The simplest energy estimates the energies of the hydrogen $L_\alpha$ and $H_\alpha$ transitions will be 6.17 GeV and 1.14 GeV, respectively. The annihilation energies of a dark electron and a dark positron are difficult to predict, but they can be near 308.5 TeV in the case of two-photon annihilation. For dark positronium (Ps- a pure Coulomb system) our energetic estimates give $L_\alpha^{Ps} \sim 3.1$ GeV and $H_\alpha^{Ps} \sim 0.58$ GeV. The hyperfine transition in the ground state ($1^3S_1 \to 1^1S_0$) has an energy of E ~ 508.3 keV. The curious possibility to explain the excess in the gamma-ray spectrum (1-3 GeV) from the Galactic center region by the $L_\alpha$ line of dark Ps arises here. The $L_\alpha$ line of ordinary Ps (2431Å) cannot be observed toward the center of our Galaxy due to strong absorption in the ultraviolet [65]. One might probably expect the development of dark-medium spectroscopy and, as a consequence, the non gravitational detection of dark matter. Dark atoms, positronium, and monopolium, can play a key role here [66, 67]. Therefore, it is important to recall the 'forgotten' magnetic charges that could form a magnetic atom, monopolium ($g^+g^-$), and even a kind of magnetic world at early cosmological epochs [67].

Artificial magnetic charges have recently been detected in laboratories [68-69], although this interpretation of them was not unambiguous. Magnetic charges (monopoles) in a free state are probable absent in the cosmos at present. However, as has already been noted, the simplest magnetic atom, monopolium, could be formed at early cosmological epochs. If two-photon annihilation energy of a light magnetic atom is ~2.4 GeV, then the energy of its ortho-para transition is ~ 282 keV. The survived (or newly formed) light (Dirac) magnetic atoms can probably be observable at the centers of galaxies and their clusters, forming a recombination –annihilation spectrum. The annihilation of heavy (t'Hooft-Polyakov) magnetic monopoles could produce ultrahigh –energy cosmic rays [67].

Note important works in Russia. In April 2015 researches from the Institute for Nuclear Research (Moscow) and Joint Institute for Nuclear Research (Dubna) as well as from a number of Russian scientific institutions entering into the Baikal collaboration deployed and put into operation a unique experimental facility, the Dubna underwater neutrino telescope at the lake Baikal. It was the first cluster of the Baikal-GVD (Gigaton Volume Detector) neutrino telescope being constructed. The Cherenkov detector is designed to investigate the high-energy neutrino flux, and, of course, the recording of heavy monopoles is also envisaged here. A review of their early results can be found in [70, 71]. We cannot but mention another important experiment, now already at the CERN; a test session of its facility was conducted in July 2016. Here, we are talking about the search for a massive photon that could emerge in the dark sector of the Universe.

## 4. Conclusion

In experiments on dark matter its dynamics must be detected $\rho_\Lambda=\rho_\Lambda(t)$, but an clear result can be seen only at high red shifts. For example, the values of dark energy at z=1 and 2 need to be compared (a predictive difference is near 4). At low red shifts (z<1), the difference in vacuum energy density will be virtually undetectable). The effects of the presence of vacuum energy can be measured in a laboratory (atomic interferometry [73]). Some of the effects related to the vacuum evolution were already discussed in [74], where a running $\Lambda$CDM – cosmology was proposed and a new look at the cosmological inflation and at huge entropy from the decay of the primeval vacuum was given.

The difficulty question about the vacuum stability arises. In [25] we introduced the vacuum stability condition in the SM. The mutual compensation of the positive and negative contributions to the vacuum energy density in the regime of super symmetry is forbidden by the stability condition. At a Higgs mass $m_H \sim 125$ GeV new physics is required for the vacuum stability up to the Planck scale [75], in which the SM vacuum must be asymptotically safe [76]. An absolutely stable vacuum could emerge if the Higgs mass were $m_H > 129$ GeV [77], and we probably live in a meta-stable vacuum. Note that the DESI (Dark Energy Spectroscopic Instrument) will be installed in the USA at Kitt Peak National Observatory in 2018 to investigate the influence of dark energy on the expansion of the Universe. Optical spectra for tens of millions of galaxies and quasars will be taken, covering the "nearby" Universe up to $10^9$ light years with 3D mapping. Modern technology already allows this to be done. The paper [78], where the current accelerated expansion of the Universe is also realized, was devoted to the dynamical relaxation of dark energy (the proof of the concept).

In the experiments on dark matter we expect the neutralino - axion (mixed) model of dark matter in the presence of some constraints to be confirmed [79]. Neutralinos and axions are probably the main components, although right-handed neutrinos must be in the dark matter, along with dark atoms. The entire enigma consists in the percentage composition of dark matter components. The probability of detecting the next fundamental level of matter is high. The composite models of the Higgs particle provide an elegant solution of the hierarchy problem. Even the composite model of a right-handed t-quark [8] was constructed to unify the gauge coupling [80]. As has already been noted, the SM is not final version of a complete description of nature and new physics is most likely required to explain the dark matter composition. The development of spectroscopy for dark (positronium, monopolium, and other) atoms should also be expected.

The identification of dark matter particles has the highest scientific priority. Dark matter particles are being searched for in dozens of experiments world-wide, from the SDEX experiment in China to the COUPP-60 experiment at the University of Chicago in Canada. These experiments are divided into experiments for the direct and indirect detection of dark matter particles. The experiment of the CoGeNT collaboration even has the (as yet unconfirmed) detection of WIMPs with a mass of ~10 GeV [81]. The masses of the particles constituting the dark matter can lie within the range from $10^{-15}$ to $10^{15}$ GeV, while the cross section for their annihilation into SM particles can occupy the range $10^{-76}$ - $10^{-41}$ $cm^2$.

Actually, the testing of gravity, whereby surprises can arise, should not be forgotten either. Clarifying the nature of the dark components in the Universe is a major scientific challenge. At present, this is some test of our understanding of high-energy physics. All experiments with both dark energy and dark matter are very costly, and only international collaborations can accomplish them at a high level. We also have several more new interesting projects: Higgs factories will be created for "purer" experiments with Higgs bosons [82]; experiments for the search of particles with a fractional charge are being prepared [83]. I repeat that the program of the laboratory of high-energy neutrino astrophysics (Baikal collaboration) also includes the search for magnetic charges (monopoles) [70]. Intriguing news came from the ATLAS and CMS collaborations: bosons with energy of 750 GeV ("Homer bosons) were detected at the LHC in the form of an excess in the two-photon channel [84]. Another news on high-energy neutrinos (37 events with E> 60 TeV and 3 events with E > $10^{15}$ eV is equality interesting. There events were detected by Ice Cube and Antares collaborations [85]. The Chinese electron-positron collider (BEPCII) set a world record in luminosity, ~$10^{33}$ cm$^{-2}$ s$^{-1}$ (see the collider news portal on particles www.interactions.org). Finally, note that the voids in the large-scale distribution of dark matter predicted by us in [47] have recently been detected in [86].